\documentclass[twocolumn,showpacs,aps,prb,amsmath,amssymb]{revtex4}

\usepackage{graphicx}
\usepackage{dcolumn}
\usepackage{bm}

\begin{document}


\title{Simple Combined Model for Nonlinear Excitations in DNA}

\author{ D.L. Hien$^1$, N.T. Nhan$^1$, V. Thanh Ngo$^{1,2}$, and N.A. Viet$^1$}

\affiliation{$^1$Institute of Physics and Electronics,\\ P. O. Box 429, Boho, Hanoi 10000, Vietnam\\
$^2$APCTP, Hogil Kim Memorial Building 5th floor, POSTECH,\\ Hyoja-dong, Namgu, Pohang 790-784, Korea}

\email{nvthanh@iop.vast.ac.vn}

\date{\today}

\begin{abstract}
We propose a new simple model for DNA denaturation bases on the pendulum model of Englander\cite{A1} and the microscopic model of Peyrard {\it et al.},\cite{A3} so called "combined model". The main parameters of our model are: the coupling constant $k$ along each strand, the mean stretching $y^\ast$ of the hydrogen bonds, the ratio of the damping constant and driven force $\gamma/F$. We show that both the length $L$ of unpaired bases and the velocity $v$ of kinks depend on not only the coupling constant $k$ but also the temperature $T$. Our results are in good agreement with previous works.
\end{abstract}

\pacs{87.10.+e, 87.14.-g, 87.15.-v, 87.83.+a}
\maketitle

\section{Introduction}
The structures of DNA molecule has been extensively studied during the last decade theoretically and experimentally. DNA molecule consists of a pair of molecules, organized as strands running start-to-end and joined by hydrogen bonds along their lengths, the diameter of the double stranded DNA is $2$~nm indeed, but the length could be much longer, usually. The separation of double helix structure of DNA is an important beginning point in the informatics
replication process of DNA that needs to reproduce the living matter. Accordingly, the problem of DNA denaturation has been the subject of extensive experimental and theoretical investigation. It is not only of practical interest due to its significance in biology but also of fundamental interest.

Englander {\it et al.}\cite{A1} first constructed a theory of a geometrical soliton by carrying the base rotations to the point that one side of the helix was given an entire extra $360^\circ$ twist. This theory is an explanation of the open states of DNA due to the nonlinear excitations along DNA.\cite{A2,A21,A10}

On the other hand, a simple lattice model (PB model) for the denaturation of the DNA double helix was proposed by Peyrard {\it et al.,}\cite{A3,A9,A4} and more advances are discussed in [\onlinecite{A16,A17,A18,A19,A20}]. They introduced the Morse interaction potential that depends on the transverse stretching of the hydrogen bonds between the complementary base pairs. These models could be used to explain the experimental observations on biological phenomena of DNA (see Refs.~\onlinecite{A5,A6}). However, recently it was argued \cite{A7,A12,A13,A14,A15} that thermodynamic characterization of the thermal fluctuations may differ from a dynamical characterization, which points to the need for a thorough understanding of the dynamical effects in this highly nonlinear and cooperative material.

Note that, in Englander model, the strands of a double helix are modeled as two chains of pendula (the bases), in which one of the strand is dynamic while other strand is fixed. By using the continuum approximation, the relative motion of the chain has been described by the sine-Gordon equation for a single pendulum in the gravitational potential. In Peyrard-Bishop model, only the out-off-phase displacements stretch the hydrogen bonds. In the continuum limit approximation and thermodynamic limit, the out-off-phase motion of the bases is described by an equation which is formally identical to the Schr\"{o}dinger equation for a particle in the Morse potential. Therefore the two above models started from the slightly similar two-line simple caricatures of DNA (see the figures \ref{fig:emodel} and \ref{fig:pbmodel} bellow) describing physical processes derivate from the same stretching of hydrogen bonds and motions of the similar bases, then both were reformulated to one-dimensional motions in some effective potentials with slightly similar equations. Because of that it is possible to combine the models.

In this paper we propose a new simple model for DNA denaturation based on the combination of two above simple models of the DNA, i.e., (E) and (PB) models. By using this combined model (EPB model), we study the temperature dependence behaviors of the nonlinear excitations in DNA.

The paper is organized as follows: Section \ref{sec:bg} gives a brief review of Englander model (\ref{sec:E}) and Peyrard-Bishop model (\ref{sec:PB}). In Sec.~\ref{sec:EPB}, we suggest the combined model by extending their works to consider the temperature dependence of the main parameters of the model. Some applications are shown in Sec.~\ref{sec:app}, we investigated the dependence of unpaired bases length and velocity of kinks on the temperature,
coupling constant, and external driven force. Concluding remarks are given in Section~\ref{sec:Conc}.

\section{Background}
\label{sec:bg}
\subsection{Englander pendulum model (E model)}
\label{sec:E}

\begin{figure}[t]
\includegraphics[width=7.5cm]{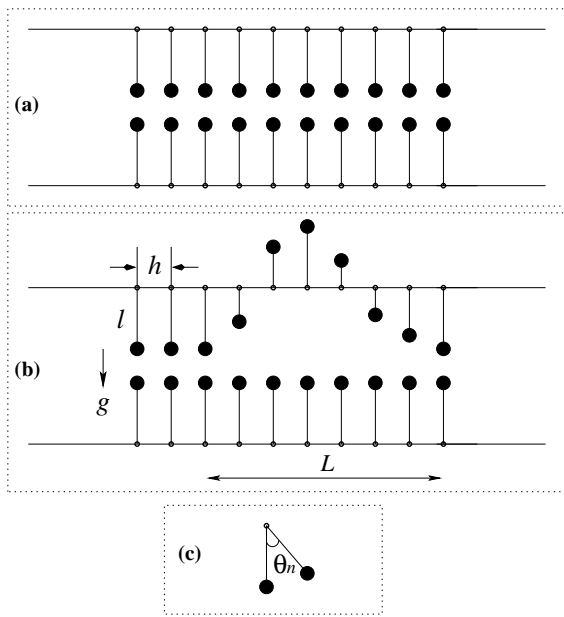}
\caption{Sketch of the Englender model.\cite{A2} (a) The ground state of the double helix. (b) Kink soliton in the sine-Gordon model, the bases of the upper chain twist through $2\pi$ over a characteristic length $L$ (the mean number of open bases) along the line. (c) The angle $\theta_n$ is the deviation of the upper base of the $n$th pair with respect to the plane defined by the fixed base pairs in the lower strand.}
\label{fig:emodel}
\end{figure}

The model of DNA proposed by Englander {\it et al.}\cite{A1} is schematically represented in Fig. \ref{fig:emodel}, where the dynamics of one of the strands of the DNA is represented as a chain of pendula; leaving the other strand fixed (the one at the bottom in Figs.~\ref{fig:emodel}a, \ref{fig:emodel}b), the base pairs of the (upper) strand behave like pendula in an gravitational field caused by the tendency of the base pairs of the two strands to be facing each other (see Fig. \ref{fig:emodel}b). It must be realized that this model describes only the dynamics of the base pairs around the sugar-phosphate backbone, which is assumed to be fixed as well.

Introducing a damping constant $\gamma$ and an external driven force $F$, these pendula are described by the discrete, dc-driven, damped sine-Gordon model.
The Hamiltonian of the coupled-pendulum system can be written as 
\begin{eqnarray}
{\mathcal H}_{E} &=&
\sum_{n}\left\{\frac{ml^2}{2}\left(\frac{\partial\theta_n}{\partial
t}\right)^2 + \frac{k}{2}(\theta_n - \theta_{n-1})^2\right. \nonumber\\
&&\left. +\ mg_n l(1 -\cos \theta_n) + \gamma\frac{\partial\theta_n}{\partial t} +
F\theta_n\right\},\label{eq:01}
\end{eqnarray}
where $m$ and $l$ is the common mass and length of pendulum, $\theta_n$ and $g_n$ is the rotation angle and the gravitational acceleration of pendulum at site $n$, respectively. $k$ is the coupling constant along the strand. 

In the mechanical unzipping or denaturation experiments, one of the DNA strands is attached to a glass bead, which is pulled by a glass micro-needle, while the other strand is attached to a glass plate, which serves as fixed reference point. The glass bead is then pulled at constant velocity, opening
or unzipping the double chain by consecutively breaking the hydrogen bonds of the base pairs, and the force used to pull it is recorded. In this context, the driven force $F$ in Eq.~(\ref{eq:01}) represents the action of the pulling on the end of one strand on the region where unzipping is taking place.\cite{A21} The damping constant $\gamma$ is introduced to fix the force scale only.\cite{A10}

In the case of homopolymer, i.e., $g_n = g$ and the pendulum spacing $h$ in the chain is small, by using the continuum approximation $\theta_n(t) \rightarrow \theta(x,t)$, we obtained the damped sine-Gordon equation
\begin{equation}
\partial^2_t\theta - \frac{k}{ml^2}\partial^2_x\theta +
\frac{g}{l}\sin\theta = -\gamma^{\ast} \partial_t\theta - F^{\ast},
\label{eq:02}
\end{equation}
here we use the notations $\gamma^{\ast} =\gamma/ml^2$ and $F^{\ast} = F/ml^2$. The soliton solutions of the kink type could be obtained by solving this equation.

The physical meaning of a kink solution in the context of DNA modeling: the bases of the upper chain perform a complete, smooth turn around the sugar-phosphate backbone, from $\theta_n=0$ to $\theta_n=2\pi$. The kink joins a sector of the chain where bases are closed, $\theta_n=0$, to another one where bases have performed a complete turn, $\theta_n=2\pi$. The chain is said to be open in its last part.

\subsection{Peyrard-Bishop microscopic model of DNA\\ (PB model)}
\label{sec:PB}

\begin{figure}[b]
\includegraphics[width=7cm]{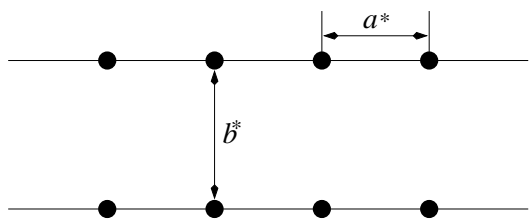}
\caption{Sketch of the PB model.}
\label{fig:pbmodel}
\end{figure}

With a particular interest in thermal stability, a microscopic model to describe the dynamics of DNA denaturation was introduced by Peyrard and Bishop in 1989.\cite{A3} In Fig. \ref{fig:pbmodel} we represent PB model of DNA, where $b^\ast$ and $a^\ast$ are the distance between two strands and the spacing between bases in a strand, respectively. The PB model has successfully reproduced the essential features of thermally induced denaturation of long DNA
chains. This model has been used to estimate the melting curves of very short heterogeneous DNA segments, in excellent quantitative agreement with experimental data.\cite{A5,A6} It provides the characteristic multistep melting of DNA sequence, and also has been used to investigate the properties of dynamic transport in DNA chains.

For each base pair, there are two degree of freedom $u_n$ and $v_n$, which correspond to the displacements of the bases from their equilibrium positions along the direction of the hydrogen bonds connect the two bases in a pair. The potential for hydrogen bonds is approximated by a Morse potential $V(r) = D[\exp(-\alpha r) - 1]^2$ with two parameters $D$ and $\alpha$. Neglecting the inhomogeneities, using a common mass $m$ for bases and the same coupling constant $k$ along each strand, the Hamiltonian for PB model reads
\begin{eqnarray}
{\mathcal H}_{PB} &=& \sum_n \left\{\frac{1}{2}\, m(\dot{u}_n^2 + \dot{v}_n^2)+V(u_n - v_n)\right. \nonumber\\
&&\left. +\ \frac{1}{2}\, k\left[(u_n - u_{n-1})^2 + (v_n - v_{n-1})^2\right]\right\}. \label{eq:04}
\end{eqnarray}
The motions of the two strands can be separated in terms of the variables $x_n = (u_n + v_n)/\sqrt{2}$, $y_n = (u_n -v_n)/\sqrt{2}$, which represent the in-phase and out-of-phase motions, respectively. Because only the out-of-phase displacements $y_n$ stretch the hydrogen bonds, the component ${\mathcal H}_y$ of Hamiltonian (\ref{eq:04}) with respect to $y$ is taken
\begin{equation}
{\mathcal H}_y = \sum_n \left\{\frac{1}{2} m\dot{y}_n^2 + \frac{k}{2} (y_n -
y_{n-1})^2 + V(\sqrt{2}y_n)\right\}\,.
\label{eq:05}
\end{equation}

As the model is assumed to be homogeneous, the result does not depend on the particular site $n$ considered. Using the transfer integral (TI) method, with the continuum limit approximation and thermodynamic limit (large number of bases pairs $N \rightarrow \infty$), the main result is dominated by the lowest eigenvalue, so the mean stretching $\left<y\right>$ of hydrogen bonds is given by
\begin{equation}
\left<y\right> = \left<\varphi_0(y)\left|y\right| \varphi_0(y)\right> = \int\varphi_0^2(y) ydy.
\label{eq:06}
\end{equation}

The mean stretching $\left<y\right>$ is interesting for the study of DNA denaturation, it increases rapidly around a particular temperature which is a characteristic of DNA denaturation. The denaturation temperature is not only sensitive to the parameters of the hydrogen bonds which bind the two strands, but also very sensitive to the intrastrand interaction constant.

\section{Combined model (EPB model)}
\label{sec:EPB}

Comparing the two above models, we see that the Englander's model explains the denaturation of DNA as the motion of soliton excitations along the chain, while the PB model has successfully describe the open states of DNA by separating the relative motion of two strands at the critical temperature. Therefore, we suggest a combined EPB model which is based on E-model with including the parameters of PB-model, i.e., taking into account the stretching $y^\ast$ of hydrogen bonds
\begin{equation}
y^\ast\ = \int\varphi_0^2(y) ydy,
\label{eq:y}
\end{equation}
where the normalized eigenfunction $\varphi_0(y)$ is obtained by solving the Schr\"{o}dinger equation for a single particle in the Morse potential
\begin{equation}
\left[-\,\frac{1}{2\beta^2 k}\, \frac{\partial^2
}{\partial y^2} + V\,(2y)\right]\varphi(y) =
\varepsilon\varphi(y), \label{eq:07}
\end{equation}
in which $\beta = 1/k_B T$ with $k_B$ being Boltzmann's constant, then we have
\begin{eqnarray}
\varphi_0(y)&=&(\sqrt{2}\alpha)^{1/2}\frac{(2 d)^{d-1/2}}{\sqrt{\Gamma(2d-1)]}}\exp\left(-de^{-\sqrt{2} \alpha y}\right)\times\nonumber\\
&&\times \exp\left[-(d-1/2)\sqrt{2} \alpha y\right].
\end{eqnarray}

Equation (\ref{eq:07}) has a discrete spectrum when $d=(\beta/\alpha)(kD)^{1/2} > 1/2$. In the case $T < T_m=2\sqrt{kD}/\alpha k_B$ the states of particle are localized, while they are delocalized for $T > T_m$. So that the critical temperature $T_m$ is considered as the melting temperature of DNA. This equation is also formally identical to the Schr\"{o}dinger equation for a quasi-particle with an effective mass $m^\ast$:
\begin{equation}
m^\ast=A\,m_0\left(\frac{T_r}{T}\right)^2,
\label{eq:mass}
\end{equation}
where $m_0$ is the mass of free electron, and $T_r=300$~K is room temperature. The dimensionless parameter $A$ is defined by
\begin{equation}
A=R_H  \frac{a_H^2k}{(k_B T_r)^2},
\label{eq:A}
\end{equation}
in which $R_H=13.6$~eV and $a_H=0.053$~nm are the Rydberg constant and Bohr radius of hydrogen atoms, respectively. Note that $A$ is proportional to the ratio of constructive and destructive energies of DNA.

\begin{figure}[b]
\includegraphics[width=7cm]{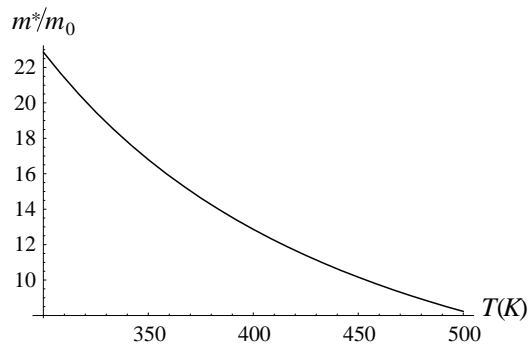}
\caption{The ratio of $m^\ast/m_0$ as a function of temperature $T$ with $k=0.2$~eV/nm$^2$.}
\label{fig:mot}
\end{figure}

\begin{figure}[h]
\includegraphics[width=7cm]{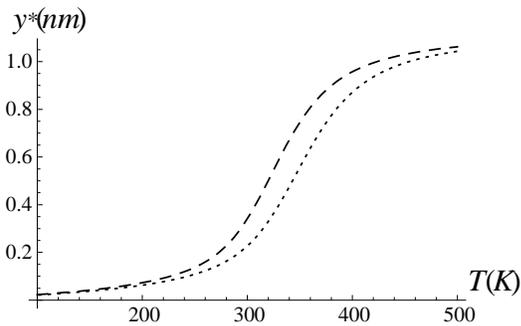}
\caption{The stretching $y^\ast$ of hydrogen bonds as a function of temperature $T$ for two values of $k$: (dashed curve) $k=0.2$~eV/nm$^2$ and (dotted curve) $k=0.23$~eV/nm$^2$.}
\label{fig:yot3k}
\end{figure}

Note that, the effective mass depends on the temperature and the structure constant of single strand (see Fig.~\ref{fig:mot}). At room temperature $T=300$~K, we found the value of effective mass $m^*\sim 22.87m_0$. We show in Fig.~\ref{fig:yot3k} the temperature dependence of the stretching $y^\ast$ of the hydrogen bonds for two values of $k$, with $D = 0.33$~eV and $\alpha=18$~nm$^{-1}$.

\begin{figure}[h]
\includegraphics[width=7cm]{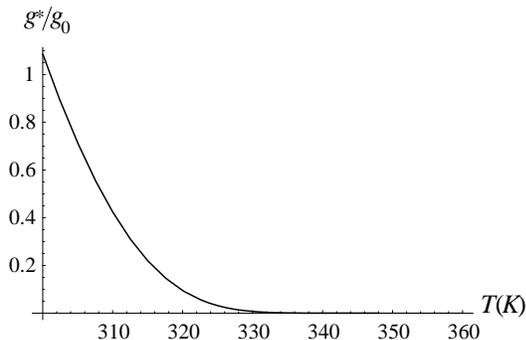}
\caption{The ratio $g^\ast/g_0$ as a function of temperature $T$ with $k=0.23$~eV/nm$^2$, $g_0=9.81$~m/s$^2$, $D = 0.33$~eV and $\alpha=18$~nm$^{-1}$.}
\label{fig:got}
\end{figure}

Now, we consider the influence of temperature on the parameters of the Englander model with the substitutions: the length and mass of the pendulum $l = y^\ast$ and $m = m^\ast$, respectively. $g=g^\ast$ with $g^\ast$ being the effective gravitational acceleration and defined by
\begin{eqnarray}
g^\ast &=&\frac{1}{m^\ast}\frac{\partial}{\partial y^\ast}V(y^\ast)\nonumber\\
 &=&-\, \frac{2D\alpha}{m^\ast}\left[\exp(-\,2\alpha y^\ast) -
\exp(-\alpha y^\ast)\right].
\label{eq:08}
\end{eqnarray}

In Fig.~\ref{fig:got} we present the ratio $g^\ast/g_0$ as a function of temperature with $g_0=9.81$~m/s$^2$. We see that $g^\ast$ decrease rapidly to zero at a critical temperature $T_m\sim 335$~K, this temperature is so-called "melting temperature" and should be discussed in next section.

It is well known that, in the absence of dissipation and force $\gamma=F=0$, Eq. (\ref{eq:02}) possesses soliton solutions of the kink type, whose expression is
\begin{equation}
\theta_{\pm}(x,t) = \pm 4\arctan\left(\exp\left[\frac{\frac{2h}{L}\ x -\sqrt{\frac{g^\ast}{y*}}\ t}{\sqrt{1-(v^\ast)^2}}\right]\right),
\label{eq:03}
\end{equation}
in which, the plus or minus sign stands for kinks or antikinks, respectively, and $0\leq v\leq 1$ represents their velocity. $L$ is a length of unpaired bases such that the mean number of open bases along the chain of pendula\cite{A1} and defined by
\begin{equation}
L = 2h(k/m^\ast g^\ast y^\ast)^{1/2}.
\label{eq:L}
\end{equation}

In the case $F=0$ and in the presence of damping, $\gamma\neq 0$, the only possible value for the velocity is $v=0$. When both damping and force are present, the balance between the two effects leads to kinks propagating at a constant, nonzero velocity. An analytical expression for that velocity can be easily derived from energy-conservation arguments (see Refs.~\onlinecite{A2,A8}),
\begin{equation}
v^\ast = v_0\left[1+ \frac{g^\ast}{y^\ast}\left(\frac{4\gamma}{\pi F}\right)^2\right]^{-1/2}.
\label{eq:V}
\end{equation}

\section{Some applications}
\label{sec:app}

We assume that the spacing of bases in a same strand does not depend on temperature, because the interaction between the bases in each strand is stronger than that between two strands. We chose the spacing $h=0.34$~nm for all numerical calculations.

\begin{figure}[b]
\includegraphics[width=7cm]{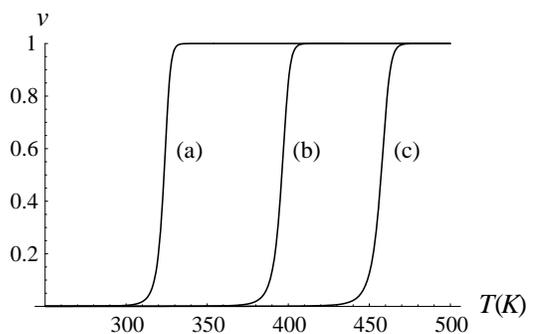}
\caption{Variation of velocity $v$ as a function of temperature $T$ with $\gamma/F=2$. (a) $k=0.2$~eV/nm$^2$, (b) $k=0.3$~eV/nm$^2$, and (c) $k=0.4$~eV/nm$^2$.}
\label{fig:vot3k}
\end{figure}

For comparing to the results of PB-model\cite{A3}, we used the following parameters: $D = 0.33$~eV, $\alpha=18$~nm$^{-1}$. By using the numerical calculations, we present in Fig.~\ref{fig:vot3k} the temperature dependence of the kink velocity for three values of coupling constant $k=0.2$~eV/nm$^2$, $k=0.3$~eV/nm$^2$, and $k=0.4$~eV/nm$^2$. The velocity $v=v^\ast/v_0$ increase with increasing the temperature and more rapidly close to $v_0=1$ at the
melting temperatures $T_m\sim 323$~K (a), $T_m\sim 395$~K (b), and $T_m\sim 458$~K (c). These results are in an excellent agreement with the results of Ref.~\onlinecite{A3} where $T_m$ is about $330$~K, $400$~K, and $460$~K for three values of $k$, respectively. The particular temperature $T_m$ is a characteristic of DNA denaturation, because the hydrogen bonds between the strands are broken at this stage.

\begin{figure}[t]
\includegraphics[width=7cm]{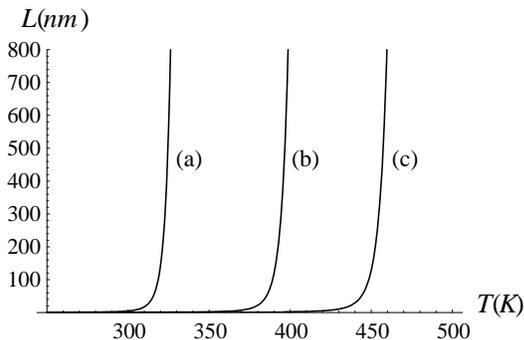}
\caption{Variation of length $L$ as a function of temperature $T$. (a) $k=0.2$~eV/nm$^2$, (b) $k=0.3$~eV/nm$^2$, and (c) $k=0.4$~eV/nm$^2$.}
\label{fig:lot3k}
\end{figure}

Figure \ref{fig:lot3k} shows the temperature dependence of the length $L$ for three values of coupling constant $k$ above. In the same manner as the kink velocity $v$, the length $L$ increases rapidly at the temperature $T\sim T_m$, so the mean number of the open bases is very large, i.e., the base-pair is said to be open at the melting temperature.

\begin{figure}[h!]
\includegraphics[width=7cm]{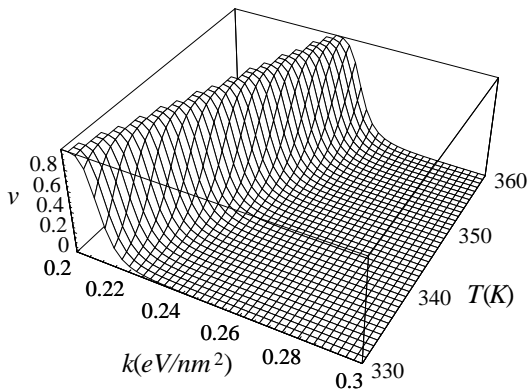}
\caption{The dependence of kink velocity on the temperature $T$ and coupling constant $k$ with $\gamma/F=2$.}
\label{fig:votk}
\end{figure}

\begin{figure}[h!]
\includegraphics[width=7cm]{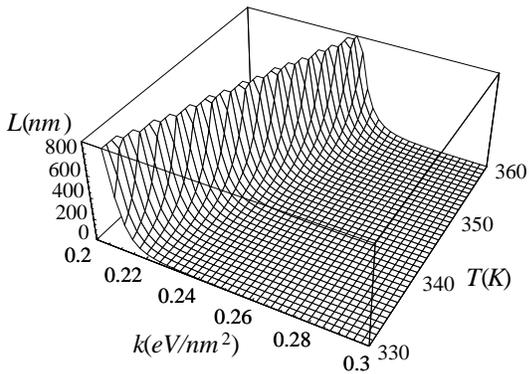}
\caption{The dependence of length $L$ on the temperature $T$ and coupling constant $k$.}
\label{fig:lotk}
\end{figure}

One of important parameters of the model is the coupling constant, as $k$ increases the denaturation temperature increases. This is consistent with the increase observed experimentally in the presence of reagents that increase the hydrophobic interactions.\cite{A11} Figures \ref{fig:votk} and \ref{fig:lotk} show that the values of $k$ must be of the order of $0.23$~eV/nm$^2$ to obtain a reasonable denaturation temperature which are obtained from the experimental results\cite{A5} $T^e_m\sim 330$~K to $345$~K. Note that our combined model works well for long DNA sequences only. It is well-known that the finite length effect of DNA sequences is important, so that our present results might not applicable for short chain of DNA, such as in [\onlinecite{A5}] with 27 base pairs and 21 base pairs sequences. The influence of finiteness length effects of DNA sequences in our model will be investigated in the future work.

\begin{figure}[h]
\includegraphics[width=7cm]{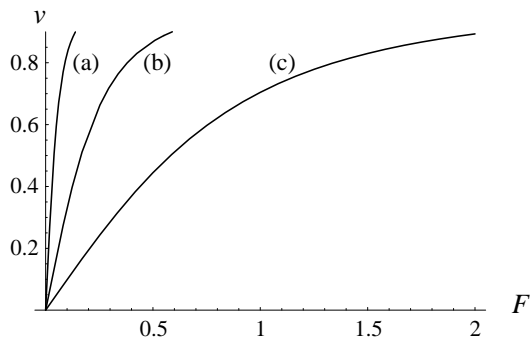}
\caption{The dependence of kink velocity on the driven force $F$ with $\gamma=0.1$, $k=0.22$~eV/nm$^2$ for three values of temperature: (a) $T=340$~K, (b) $T=335$~K, and (c) $T=330$~K.}
\label{fig:voft}
\end{figure}

Finally, we show in Fig.~\ref{fig:voft} the dependence of kink velocity on the driven force $F$ for three values of temperatures: (a) $T=340$~K, (b) $T=335$~K, and (c) $T=330$~K. The velocity increases with increasing the driven force and more rapidly at high temperature. The melting temperature could be controlled by varying the external driven force $F$. 

\section{Conclusions}
\label{sec:Conc}

In this work we suggest a combined EPB model to describe the dynamics of DNA denaturation, which is based on Englander's model with parameters derived from microscopic PB-model. Using this combined model (EPB model) with the motion of bases in both $x$ and $y$ directions, we investigated the temperature dependence behaviors of the nonlinear excitations in DNA. We also studied the dependence of reasonable denaturation temperature on the coupling constant of
bases and the driving force. The temperature dependence of the kinks velocity and the length of unpaired bases are shown and discussed.

This simple model could be applied to study other physical quantities of E-model such as the effects of inhomogeneity arising from the genetic sequence and the influence of temperature on the kink propagation. These are objects of our future investigations.

\end{document}